\documentclass[10pt,conference]{IEEEtran}

\usepackage{graphicx,cite,hhline,amsmath,amssymb,psfrag,cite,amsfonts,color,stfloats}
\usepackage{epsfig}
\usepackage{epstopdf}
\usepackage{acronym}
\usepackage[table]{xcolor}
\usepackage{bm}
\usepackage{arydshln}
\usepackage{tikz}
\usepackage[utf8]{inputenc}
\usepackage{pgfplots} 
\usepackage{pgfgantt}
\usepackage{pdflscape}
\pgfplotsset{compat=newest} 
\pgfplotsset{plot coordinates/math parser=false}
\usepackage{graphicx}
\usepackage{subcaption}

\newlength\figureheight 
\newlength\figurewidth

\DeclareMathOperator*{\argmin}{arg\,min}
\newcommand{\dij} {d_{ij}^{(k)}}

\newcommand{\sdo} { \sigma_{0, \mathsf{r}} }
\newcommand{\sbo} { \sigma_{0,\mathsf{b}} }

\acrodef{AV}{autonomous vehicle}
\acrodef{AOA}{angle-of-arrival}
\acrodef{CRLB}{Cram\'er-Rao lower bound}
\acrodef{FANET}{flying ad-hoc network} 
\acrodef{FIM}{Fisher Information Matrix}
\acrodef{GPS}{Global Positioning System}
\acrodef{LOS}{line-of-sight}
\acrodef{MILP}{Mixed Integer Linear Programming}
\acrodef{MCP}{Model Predictive Control}
\acrodef{NLOS}{non line-of-sight}
\acrodef{NFER} {Near-Field Electromagnetic Ranging}
\acrodef{PEB}{Position Error Bound}
\acrodef{RV}{random variable}
\acrodef{RSS}{received signal strength}
\acrodef{RMSE}{root mean squared error}
\acrodef{SLAM}{Simultaneous Localization and Mapping}
\acrodef{std}{standard deviation}
\acrodef{SNR}{signal-to-noise ratio}
\acrodef{TOA}{time-of-arrival}
\acrodef{UAV}{unmanned aerial vehicle}
\acrodef{UWB}{ultrawide bandwidth}
\acrodef{WSN}{wireless sensor network}

\acrodef{UGV}{unmanned ground vehicle}
\acrodef{AUV}{autonomous underwater vehicle}

\begin{document}

\title{Joint Indoor Localization and Navigation of UAVs \\ for Network Formation Control 
\vspace{-0.5cm}
}

\author{
    \IEEEauthorblockN{Anna Guerra\IEEEauthorrefmark{1}, Davide Dardari\IEEEauthorrefmark{1}, and ~Petar M. Djuri\'c\IEEEauthorrefmark{2}}\\
    \IEEEauthorblockA{\IEEEauthorrefmark{1}DEI-CNIT, University of Bologna,
Via dell'Universit\'a, 47522 Cesena, Italy. E-mail: \{anna.guerra3, davide.dardari\}@unibo.it}
      \IEEEauthorblockA{\IEEEauthorrefmark{2} ECE, Stony Brook University, Stony Brook, NY 11794, USA. E-mail: petar.djuric@stonybrook.edu}
}

\maketitle
\bstctlcite{IEEEexample:BSTcontrol}
\pagenumbering{gobble}

\begin{abstract} 
In this paper, we propose a joint indoor localization and navigation algorithm to enable a swarm of unmanned aerial vehicles (UAVs) to deploy in a specific spatial formation in indoor environments. In the envisioned scenario, we consider a static user acting as a central unit whose main task is to acquire all the UAV measurements carrying position-dependent information and to estimate the UAV positions when there is no existing infrastructure for positioning. Subsequently, the user exploits the estimated positions as inputs for the navigation control with the aim of  deploying the UAVs in a desired formation in space (\textit{formation shaping}). The user plans  the trajectory of each UAV in real time, guaranteeing a safe navigation in the presence of obstacles. The proposed algorithm guides the UAVs to their desired final locations with good accuracy.
\end{abstract}


\IEEEpeerreviewmaketitle

\section{Introduction}
\label{sec:introduction}
In the last couple of decades, multi-agent networks have often been studied  especially in the robotic and control research fields \cite{liu2018survey, oh2015survey}.
Initially, the interest was generated by   military applications because autonomous agents, able to operate without a pilot, represent a valid alternative to human soldiers in high-risk missions  \cite{sullivan2006evolution}. In this context, trajectory optimization and motion control have been analyzed for  \acp{UAV}, \acp{UGV}, and \acp{AUV}, together with their capability of formation shaping and maintenance \cite{egerstedt2001formation,stone2000multiagent}.
Another broad domain of applications has been in space exploration, where the autonomy and flexibility of the agents play  crucial roles \cite{roehr2014reconfigurable}.

Recently, the idea of using swarms or fleets of \acp{AV} with hundreds of agents has been put forth because of the necessity of enhancing the robustness in completing assigned tasks to \acp{AV}. 
Further, such fleets permit to allocate different tasks to different sub-swarms: for example, the encirclement of an obstacle by a group of \acp{AV} and the arrival at the base by another \cite{sharma2009collision}. This work has opened  
research on intelligent agents for
outdoor civil applications, such as delivery, logistic, precision agriculture, emergency or post-disaster events \cite{balakirsky2007towards,liu2016multirobot,bayerlein2018trajectory}. 

In our previous work \cite{guerra2018collaborative}, the \acp{UAV}  fly outdoors and their mission is to navigate in a way that maximizes the capability of localizing a target. 
Going a step forward, one could easily imagine a setting where 
the \acp{UAV} enter a building with  harsh propagation conditions, with many obstacles and a drastically reduced space for maneuvering. 
Indoor environments provide 
many challenges from scientific and engineering  points of view because the \acp{UAV} do not have access for localization by a global reference system, such as by satellites (GPS) or by an ad-hoc positioning infrastructure (anchor nodes). Therefore, in addition to their traditional navigation tasks, they must perform anchor-free localization based on inter-UAV relative measurements \cite{patwari2005locating}.
This, in turn, will require superior sensing and communication capabilities than those needed in outdoor applications. 

In an anchor-free scenario, the \acp{UAV} have to measure some inter-UAV position-dependent quantities, such as distances and angles, and then, communicate them to other UAVs or to a central node (if it exists). The latter will be responsible for processing the received data and for inferring the positions of all the UAVs of the network. In this regard, several anchor-free localization algorithms have been proposed in the literature, as, for example, the ones described in \cite{wymeersch2009cooperative} and the references therein.
 
While addressing the problems of localization and navigation,  physical and dynamical constraints must be taken into account. The \acp{UAV} are expected to be lightweight, the hardware on board cannot be complex (low complexity), and the necessary computations have to be fast. For example, the interval between the instant in which the current position is estimated and the next position is reached should be kept as short as possible (low latency). Obviously, satisfying these requirements may come at the expense of a lower localization and formation accuracy, and, thus, it is important to analyze the trade-off between the technological issues and the attainable performance.
\begin{figure}[t!]
  \centering
\includegraphics[width=0.45\textwidth]{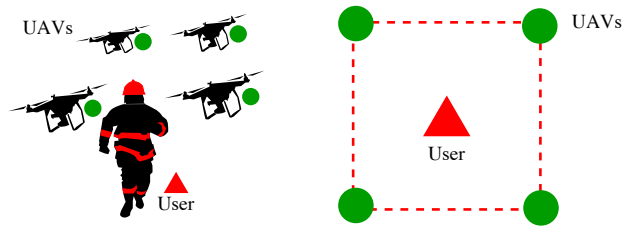}
 \caption{An indoor application scenario using a user-centric network of \acp{UAV}.}
\label{fig:application}
\end{figure}
%

In this paper, we study a network of multiple \acp{UAV} and a central user that can either be a drone, a mobile device, or a terrestrial vehicle, operating in indoor environments (see Fig.~\ref{fig:application}). The goal of the network is twofold: first, the \acp{UAV} should be localized with respect to the user (infrastructure-less localization) and, subsequently, they should deploy according to a desired topology (formation shape) and guidance by the user. The system is centralized: 
all the computational load is handled by the user, and in this way the \acp{UAV} can save their energy and complete parallel tasks while waiting for the navigation commands. Further, we suppose that all the \acp{UAV} are able of exchanging distance or \ac{AOA} measurements with all the other \acp{UAV} in the network and also with the user. The accuracy of such measurements clearly depends on the technology available on board.
For example, for distance estimation, the \ac{UWB} technology provides higher performance with respect to what can be obtained by \ac{RSS} measurements \cite{jourdan2008position,gezici2008survey}. Likewise,  the availability of antenna array measurements permits to achieve much more precise \ac{AOA} estimates than those obtained from power observations. In \cite{petitjean2018pimrc,isaacs2014quadrotor,isaacs2014gps}, it is shown how \ac{AOA} information can be extrapolated from \ac{RSS} data by exploiting the possibility of pointing the UAV's sensor antenna in different spatial directions using UAV rotations and by searching for the \ac{RSS} peak. 
Here, we chose to work with simple  
sensors and studied the achievable localization and formation accuracy based on their measurements. 

The paper is organized as follows. Section \ref{sec:problemstate} illustrates the problem, Sec.~\ref{sec:relativeCRB} describes the characterization of the relative localization error, Sec.~\ref{sec:controllaw} presents the approach used for the control of the \acp{UAV}, Sec.~\ref{sec:cs} describes the results and, finally, Sec.~\ref{sec:conclusions} draws conclusions and discusses possible future extensions.

\section{Problem Statement}
\label{sec:problemstate}
In this paper, we aim at studying how to navigate \acp{UAV} in forming a network with a desired topology with respect to a static user in indoor environments where GPS signals are not available and in the absence of an ad-hoc positioning infrastructure. A possible application is the deployment of an autonomous team of \acp{UAV} acting as radio ``flying eyes" and helping a rescuer (user) in completing his/her job, for example, by enhancing his/her capability for sensing the surrounding environment, see Fig.~\ref{fig:application}.

\begin{figure}[t!]
\psfrag{N}[c][c][0.75]{\qquad \quad\quad \quad $\mathsf{T2}:$ {Navigation}} %
\psfrag{L}[c][c][0.75]{\qquad\quad\,\,\, {Localization}} 
\psfrag{R}[c][c][0.75]{\qquad\qquad $\mathsf{T1}:$ {Relative}} 
\psfrag{C}[c][c][0.75]{\qquad \quad Cost Fun.}
\psfrag{F}[c][c][0.75]{\qquad\quad\, Derivation}
\psfrag{Nav}[c][c][0.7]{Navigation}
\psfrag{Cos}[c][c][0.7]{Constraints}
\psfrag{Opt}[lc][lc][0.75]{Optimiz.}
\psfrag{Solv}[lc][lc][0.75]{Solver}
\psfrag{Contr}[lc][lc][0.75]{Control}
\psfrag{Law}[lc][lc][0.75]{Law}
\psfrag{p}[c][c][0.6]{$\hat{\bm{\theta}}^{(k)}$}
\psfrag{b}[c][c][0.6]{$\mathbf{b}^{(k)}$}
\psfrag{ps}[c][c][0.6]{$\bm{\theta}^*$ Desired Topology}
\psfrag{z}[lc][lc][0.6]{${\mathbf{z}}^{(k)}$}
\psfrag{u}[c][c][0.6]{$\,\,\,{\mathbf{u}}^{(k+1)}$}
\psfrag{d}[c][c][0.6]{$\lVert \hat{\bm{\theta}}^{(k+1)} - \bm{\theta}^{*} \rVert_2$}
\psfrag{n}[c][c][0.6]{$\nabla_{\hat{\mathbf{p}}_{i}^{(k)}} \, \left( d_i^* \right)$}
\psfrag{T}[lc][lc][0.8]{User}
  \centering
\includegraphics[width=0.45\textwidth]{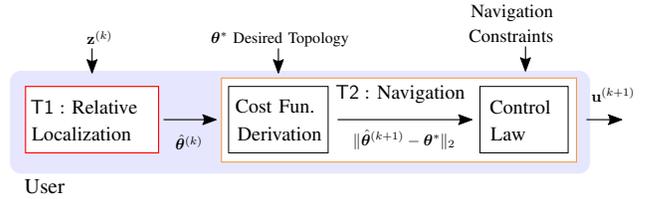}
 \caption{The user's block-diagram for joint localization and navigation.}
\label{fig:scheme}
\end{figure}

The network is composed of a single user and $N$ \acp{UAV}. The \acp{UAV} are equipped with radio sensors and exchange ranging or bearing (\ac{AOA}) measurements with each other and with the user. The user acts as the center of the network and as a (centralized) processing unit. More specifically, it performs two high-level tasks: the first is the relative localization of all the \acp{UAV} that compose the network with respect to itself (namely $\mathsf{T1}$ in Fig.~\ref{fig:scheme}); and the second is the computation of  control signals for \ac{UAV} navigation (namely $\mathsf{T2}$ in Fig.~\ref{fig:scheme}). To do this, the user collects all the measurements from the \acp{UAV} and estimates their relative 
positions with respect to  
an arbitrary selected coordinate system. 
Then, it evaluates the distance to be travelled by each \ac{UAV} to attain the desired formation and sends back the navigation commands to each \ac{UAV} at each time instant. 

To formalize the problem, we indicate with 
\begin{equation}
{\bm{\theta}}^{(k)}=\left[\left({\mathbf{x}}^{(k)} \right)^\mathsf{T}, \left({\mathbf{y}}^{(k)}\right)^\mathsf{T} \right]^\mathsf{T} 
\end{equation}
the actual position of all the \acp{UAV} composing the network at time instant $k$  (state vector), where  \begin{align}
\mathbf{x}^{(k)}&=\left[{x}_1^{(k)}, \ldots, {x}_i^{(k)}, \ldots, {x}_N^{(k)} \right]^\mathsf{T},\\ \mathbf{y}^{(k)} &= \left[{y}_1^{(k)}, \ldots, {y}_i^{(k)}, \ldots, {y}_N^{(k)} \right]^\mathsf{T}
\end{align}
are their $x$ and $y$ Cartesian coordinates, respectively. For notation simplicity, we consider a two-dimensional scenario -- the extension to a three-dimensional setting is straightforward. 
We assume that the user is located at the origin of the coordinate system (defined below), i.e., ${\mathbf{p}}_0^{(k)}=\left[ x_0^{(k)}, y_0^{(k)} \right]^\mathsf{T}=\left[ 0,0 \right]^\mathsf{T}$. 

The locations of the \acp{UAV} with respect to the user are unknown. Based on the exchanged measurements between the UAVs, such positions can be inferred by considering a coordinate system with the $x$-axis defined by a baseline centered at the user position, i.e.,  $\mathbf{b}^{(k)}=\left[ \mathbf{p}_{0}^{(k)}, \hat{\mathbf{p}}_1^{(k)} \right]^\mathsf{T}$ with $ \hat{\mathbf{p}}_1^{(k)}=\left[\hat{d}_1^{(k)}, 0 \right]^\mathsf{T}$ being the position of the first UAV. The $y$-axis is obtained by rotating the $x$-axis counterclockwise by an angle of $\pi/2$. 

In the sequel, we denote with $N_{\mathsf{r}}$ and $N_{\mathsf{u}}$ the number of UAVs whose positions are known and unknown, respectively. In our case,  $N_{\mathsf{r}}=1$ and $N_{\mathsf{u}}=N-1$.
The user estimates the corresponding $2\,N_{\mathsf{u}}$ unknown parameters and stacks their values in
\begin{equation}\label{eq:pos_est}
\hat{\bm{\theta}}^{(k)}=\left[\left( \hat{\mathbf{x}}^{(k)} \right)^\mathsf{T}, \left(\hat{\mathbf{y}}^{(k)}\right)^\mathsf{T} \right]^\mathsf{T}\, = {\bm{\theta}}^{(k)} + \bm{\omega}^{(k)},
\end{equation}
where $\hat{\mathbf{x}}^{(k)}$ and $\hat{\mathbf{y}}^{(k)}$ are the vectors with the estimated $x$ and $y$ coordinates, and $\bm{\omega}^{(k)}$ is the position estimation error. 

Once the \ac{UAV} positions are estimated, the user provides the control signals to attain the desired configuration of the \acp{UAV} with respect to the user, i.e., ${\bm{\theta}}^{*}=\left[\left({\mathbf{x}}^{*} \right)^\mathsf{T}, \left({\mathbf{y}}^{*}\right)^\mathsf{T} \right]^\mathsf{T}$, which does not depend on 
time. 

Finally, the formation problem can be formulated as a minimization problem, that is, 
\begin{equation}\label{eq:problem}
\begin{aligned}
\left({\bm{\theta}}^{(k+1)}\right)^{\star}= &\argmin_{\hat{\bm{\theta}}^{(k+1)} \in \mathbb{R}^{2N}}  \,\,  \lVert \hat{\bm{\theta}}^{(k+1)} - \bm{\theta}^{*} \rVert, \\
&\text{\,\,subject to}
\quad  d_{ij}^{(k+1)} \geq d_{\mathsf{S} }\\
&\qquad\qquad\quad\,\, \mathcal{T} \cap \mathcal{O} =  \varnothing  \\
&\qquad\qquad\quad\,\, \lVert \hat{\bm{\theta}}_i^{(k+1)}-\hat{\bm{\theta}}_i^{(k)} \rVert = v
\end{aligned}
\end{equation}
%
for $i = 0, \ldots, N$, $j\neq i$, where $d_{ij}^{(k)}$ is the distance between the $i$th and $j$th UAV at time instant $k$ as perceived by the on-board proximity sensors, and $d_{\mathsf{S}} $  is the inter-UAV safety distance. Moreover, $\mathcal{T}$ represents the set of trajectories followed by all the \acp{UAV} and $\mathcal{O}=\bigcup_{i=1}^{N_\text{obs}} \mathcal{O}_i$ the set of all obstacles  from which the \acp{UAV} should keep a safety distance equal to $d_{\mathsf{O}}$. Finally, the last constraint imposes a constant \ac{UAV} speed equal to $v$.
Then, the control signal to be applied at the $i$th \ac{UAV} to satisfy \eqref{eq:problem} is 
\begin{equation}\label{eq:controlsignal}
\begin{aligned}
 \mathbf{u}^{(k+1)}=\left(\bm{\theta}^{(k+1)}\right)^{\star}-\hat{\bm{\theta}}_i^{(k)},
\end{aligned}
\end{equation}
where its $i$th element $\mathbf{u}_i^{(k+1)}=\left[ \Delta x_i^{(k+1)},\,  \Delta y_i^{(k+1)}\right]^{\mathsf{T}}$ is the spatial step to be taken by the $i$th \ac{UAV} to lower the distance toward the final desired position.

\section{Infrastructure-less Localization}
\label{sec:relativeCRB}

In this section, we characterize the position estimation error in \eqref{eq:pos_est}. In general, the relative localization uncertainty depends on the implemented  estimator. 
To gain an insight about the best attainable performance, we consider the error having zero-mean Gaussian distribution with a covariance matrix $\bm{\kappa}^{(k)}$  equal to the relative \ac{CRLB} of the UAV positions. In the following we  derive the log-likelihood function starting from the specific observation model and, then, we derive the \ac{CRLB}.

\subsection{Observation model}
\label{sec:obsmodel}

We denote with $\mathbf{z}^{(k)}$ the vector containing all the measurements available at the user at time instant $k$,
\begin{equation} 
\text{$\mathbf{z}^{(k)}$}=\left[   \ldots , z_{ij}^{(k)} , \ldots  \right]^\mathsf{T},
\end{equation}
%
with $i,j=0, \ldots, N$ and where we consider the measurement between the $(i,j)$th pair of UAVs only once to not burden the processing of measurements, i.e., $j>i$.
The generic element is the measurement between the $i$th and $j$th UAV  given by
\begin{align}\label{eq:interUAVmeas}
&z_{ij}^{(k)}=  \\
&\begin{cases}
h_{ij}^{(k)}\left( \bm{\theta}^{(k)} \right) + v_{ij}^{(k)}, \qquad\qquad\qquad\qquad\qquad\quad\,\,\,\, \mathsf{ranging},\\
p_{ij}^{(k)}\, \left( h_{ij}^{(k)}\left( \bm{\theta}^{(k)}  \right) + v_{ij}^{(k)}  \right) +(1-p_{ij}^{(k)})\, q_{ij}^{(k)}, \quad\,\, \mathsf{bearing},
\end{cases} \nonumber
\end{align}
where $p_{ij}^{(k)}\!\!=\!\!\left\{1,0 \right\}\!\!=\!\!\left\{\text{LOS},\text{NLOS} \right\}$ indicates the presence or absence of a \ac{NLOS} propagation condition between these \acp{UAV}, $q_{ij}^{(k)} \!\!\!\sim\mathcal{U}\left(0, 2\pi \right)$ models the presence of outliers due to multipath between them, and $h_{ij}^{(k)}\left( \bm{\theta}^{(k)}  \right)$ is a function of the state vector, where
\begin{align}
 &h_{ij}^{(k)}\left( \bm{\theta}^{(k)} \right)=   \\
 &\begin{cases}
 d_{ij}^{(k)}= \sqrt{ \left(\Delta x_{ij}^{(k)} \right)^2 + \left(\Delta y_{ij}^{(k)} \right)^2}, \,\,\,\,\,\,\,\qquad\qquad\,\,\, \mathsf{ranging},\\
 \phi_{ij}^{(k)}= \arctan\left( \frac{\Delta y_{ij}^{(k)}}{\Delta x_{ij}^{(k)}} \right), \qquad\qquad\qquad\qquad\,\,\,\,\,\,\,  \mathsf{bearing},
 \end{cases} \nonumber
\end{align}
with $d_{ij}^{(k)}$ and $\phi_{ij}^{(k)}$ being the distance and the angle between the $i$th and $j$th UAV at time instant $k$, with $\Delta x_{ij}^{(k)} =x_i^{(k)}- x_j^{(k)}$ and $\Delta y_{ij}^{(k)} =y_i^{(k)}- y_j^{(k)}$. The measurement error is indicated with $v_{ij}^{(k)}$ modeled as a zero-mean Gaussian random variable with a standard deviation given by
\begin{equation}\label{eq:sigma_meas}
\sigma_{ij}^{(k)} =
\begin{cases}
\sigma_{0,\mathsf{r}}  \cdot d_{ij}^{(k)}\left( \bm{\theta}^{(k)} \right), \qquad\qquad\qquad\qquad\quad\,\, \mathsf{ranging},\\
 \sigma_{0,\mathsf{b}}, \qquad\qquad\qquad\qquad\qquad\qquad\qquad\,\,\, \mathsf{bearing},
\end{cases}
\end{equation}
where the ranging model is distance dependent, with $\sdo$ being the ranging standard deviation at the reference distance (i.e., $d_0=1\,$m) and the bearing model is considered constant with respect to the state vector. For \ac{RSS}-based ranging observations, it is possible to model the ranging standard deviation as $\sdo=\frac{\ln 10}{10 \, \alpha} \left(\sigma_{\mathsf{sh}}+\left(1-p_{ij}^{(k)}\right)\, \sigma_{\mathsf{b}} \right)$, where  $\alpha$ is the path-loss exponent, $\sigma_{\mathsf{sh}}$ corresponds to shadowing in dB and $\sigma_{\mathsf{b}}$ represents a bias term due to the \ac{NLOS} condition.

Consequently, we can write the likelihood function as $f\left(\mathbf{z}^{(k)}  \lvert \bm{\theta}^{(k)}\right)= \mathcal{N} \left(\mathbf{z}^{(k)};\mathbf{h}^{(k)}, \mathbf{Q}^{(k)}  \right)$ where $\mathbf{h}^{(k)}=\left[\ldots, {h}_{ij}^{(k)},\ldots \right]^{\mathsf{T}}$ contains the expected ranging and bearing values and  $\mathbf{Q}^{(k)}=\text{diag}\left(\ldots, \left(\sigma^{(k)}_{ij}\right)^2, \ldots \right)$ is the diagonal covariance matrix. Given this observation model, the log-likelihood function {available} at the user at the $k$th time slot is 
\begin{align}\label{eq:loglik}
\Lambda\left(\mathbf{z}^{(k)}  \lvert \bm{\theta}^{(k)} \right) &= \sum_{i=0}^{N} \sum_{\substack{j=0 \\ j>i}}^{N} \ln f\left(z_{ij}^{(k)} \lvert \bm{\theta}^{(k)} \right).
\end{align}

In the literature, different localization algorithms have been proposed for dealing with the absence of an ad-hoc positioning infrastructure. To name a few, an anchor-free localization using hop count (AFL) has been proposed in \cite{priyantha2003anchor}, maximum likelihood and least square schemes are described in \cite{moses2003self}, and Bayesian schemes are investigated in \cite{wymeersch2009cooperative}. Surveys on infrastructure-less positioning can be found in \cite{wymeersch2009cooperative,patwari2005locating}. 

In the rest of the paper, we consider an unbiased estimator with a covariance matrix defined by the relative \ac{CRLB}, derived in the following section. More specifically, we model the position estimates as in \eqref{eq:pos_est} with $\bm{\omega}^{(k)} \sim \mathcal{N}\left(\mathbf{0}, \bm{\kappa}^{(k)} \right)$, and $\bm{\kappa}^{(k)}=\bm{\kappa}\!\left(\hat{{\bm{\theta}}}^{(k-1)} \right)$ being the corresponding \ac{CRLB} evaluated according to the previous position estimates. This choice allows for investigation of the formation performance in a more general way without dealing with a specific localization algorithm, and  for obtaining the maximum achievable accuracy.

\subsection{Anchor-free \ac{CRLB}}

In this section, we evaluate the anchor-free \ac{CRLB} that will be used to model the covariance matrix of the position estimates in \eqref{eq:pos_est}.
The performance of any unbiased estimator $\hat{\bm{\theta}}^{(k)}$ can be bounded by the \ac{CRLB}, namely $\bm{\kappa}\!\left({\bm{\theta}^{(k)}} \right)$, defined as \cite{van2004detection}
\begin{equation}
\mathbb{E}\left\{\!\left[\hat{\bm{\theta}}^{(k)}-{\bm{\theta}^{(k)}} \right]\!\! \left[\hat{\bm{\theta}}^{(k)}-{\bm{\theta}^{(k)}}\right]^\mathsf{T} \!\right\}\!\! \succeq \!\mathbf{J}^{-1}\!\left( {\bm{\theta}^{(k)}}  \right)=\bm{\kappa}\!\left({\bm{\theta}^{(k)}} \right),
\end{equation}
where $\mathbf{A} \succeq \mathbf{B}$ means that $\mathbf{A}-\mathbf{B}$ is positive semi-definite, and   $\mathbf{J}\!\left( {\bm{\theta}^{(k)}}  \right)$ is the $2\, N_{\mathsf{u}} \times 2\, N_{\mathsf{u}}$ \ac{FIM}, i.e., 
\begin{align}\label{eq:classicalFIM}
\mathbf{J}\!\left( \bm{\theta}^{(k)} \right) \!\!&= \mathbb{E}_{\mathbf{z}^{(k)} } \!\! \left\{ \left[ \nabla_{\bm{\theta}^{(k)}} \, \Lambda\left(\mathbf{z}^{(k)}  \lvert \bm{\theta}^{(k)} \right) \right]\!\! \left[ \nabla_{\bm{\theta}^{(k)}} \, \Lambda \left( \mathbf{z}^{(k)}  \lvert \bm{\theta}^{(k)} \right) \right]^\mathsf{T} \right\} \nonumber \\
&=\left[\begin{array}{cc} \mathbf{J}_{\mathsf{xx}}^{(k)} & \mathbf{J}_{\mathsf{xy}}^{(k)} \\ \left(\mathbf{J}_{\mathsf{xy}}^{(k)}\right)^\mathsf{T} & \mathbf{J}_{\mathsf{yy}}^{(k)} \end{array}  \right],
\end{align}
\noindent with the subscripts $\mathsf{x}$ and $\mathsf{y}$ indicating the Cartesian position coordinates of the \acp{UAV} and where the generic elements of the sub-\acp{FIM} are \cite{patwari2005locating}
\begin{align}\label{eq:FIMe}
& \left[\mathbf{J}_{\mathsf{xx}}^{(k)} \right]_{mn} \!\!\!\!= \begin{cases} \sum_{i}\, A_{mi}^{(k)}\, {\left( \Delta x_{mi}^{(k)} \right)^2}/{\left(d_{mi}^{(k)}\right)^{s}}, \,\,\,\,\,\, \,\,\,\,  \text{  $m=n$},\\  
- A_{mn}^{(k)}\, {\left( \Delta x_{mn}^{(k)} \right)^2}/{\left(d_{mn}^{(k)}\right)^{s}}, \,\,\,\,\,\,\,\,\, \,\,\,\, \text{  $m \neq n$,} \end{cases} \nonumber \\
& \left[\mathbf{J}_{\mathsf{xy}}^{(k)} \right]_{mn} \!\!\!\!=\begin{cases} \sum_{i}\, A_{mi}^{(k)}\, { \Delta x_{mi}^{(k)}\, \Delta y_{mi}^{(k)}}/{\left(d_{mi}^{(k)}\right)^{s}}, \,\,\,\, \text{  $m=n$,}\\  
- A_{mn}^{(k)}\, {\Delta x_{mn}^{(k)}\, \Delta y_{mn}^{(k)}}/{\left(d_{mn}^{(k)}\right)^{s}}, \,\,\,\,\,\,\, \text{  $m \neq n$,} \end{cases}  \nonumber\\
& \left[\mathbf{J}_{\mathsf{yy}}^{(k)} \right]_{mn} \!\!\!\!=  \begin{cases} \sum_{i}\, A_{mi}^{(k)}\, {\left( \Delta y_{mi}^{(k)} \right)^2}/{\left(d_{mi}^{(k)}\right)^{s}}, \,\,\,\,\,\,\,\,\,\,\, \text{  $m=n$,}\\  
- A_{mn}^{(k)}\, {\left( \Delta y_{mn}^{(k)} \right)^2}/{\left(d_{mn}^{(k)}\right)^{s}}, \,\,\,\,\,\,\,\,\,\,\,\,\,\, \text{  $m \neq n$,} \end{cases} \nonumber\\
\end{align}
where $i=0, \ldots, N$ and $m=n=1, \ldots, N_{\mathsf{u}}$. The exponent $s$ is equal to $2$ in the case of ranging measurements, and $s=4$ in the case of bearing. The  coefficient $A_{mi}^{(k)}$ is defined by 
\begin{align}\label{eq:A}
&A_{mi}^{(k)}=\begin{cases} 
 (1+ 2\, \xi \, \sdo )/\left(\sigma_{mi}^{(k)}\right)^2, \qquad \mathsf{ranging},\\
 {1}/\left(\sbo^2\right), \qquad \qquad \qquad\quad\,\,  \mathsf{bearing},
  \end{cases}
\end{align}
where $\xi=0$ when there is a model mismatch, i.e., when the dependence of $\sigma_{mi}^{(k)}$ from the state vector $\bm{\theta}^{(k)}$ in \eqref{eq:sigma_meas} is not considered in the \ac{FIM} evaluation; otherwise $\xi=1$.

\section{The Control Law}
\label{sec:controllaw}

The constrained minimization problem in \eqref{eq:problem} can be solved using the \textit{projection gradient} method  \cite{luenberger1984linear}. Then the control signal of the $i$th UAV is given by
\begin{equation}\label{eq:update_g2}
\mathbf{u}_i^{(k+1)} \!=\! -\gamma\, \mathbf{P}\, \nabla_{\hat{\bm{\theta}}_{i}^{(k)}} \, \left(\lVert \hat{\bm{\theta}}_i^{(k)} - \bm{\theta}_i^{*} \rVert \right) \!-\! \mathbf{N}\left(\mathbf{N}^\mathsf{T} \mathbf{N}\right)^{-1}\mathbf{g},
\end{equation}
where $\gamma$ represents the spatial step, and {$\nabla_{\bm{\hat{\theta}}_{i}^{(k)}} \, \left(\cdot \right)$} is the gradient operator with respect to the {UAV}'s estimated position, and it is
\begin{align}
&\nabla_{\hat{{x}}_{i}^{(k)}} \, \left(\lVert \hat{\bm{\theta}}_i^{(k)} - \bm{\theta}_i^{*} \rVert\right)=\cos\left(\hat{\alpha}_i^{(k)} \right), \\
&\nabla_{\hat{{y}}_{i}^{(k)}} \, \left(\lVert \hat{\bm{\theta}}_i^{(k)} - \bm{\theta}_i^{*} \rVert\right)=\sin\left(\hat{\alpha}_i^{(k)} \right),
\end{align}
where $\hat{\alpha}_i^{(k)}$ is the angle between the estimated position of the  $i$th \ac{UAV} and its desired position at time instant $k$.
\noindent The projection matrix is denoted with $\mathbf{P}=\mathbf{I}-\mathbf{N}\left(\mathbf{N}^\mathsf{T} \mathbf{N}\right)^{-1}\mathbf{N}^\mathsf{T}$, where $\mathbf{I}$ is the identity matrix and $\mathbf{N}=\nabla_{\hat{\bm{\theta}}_{i}^{(k)}}\left(\mathbf{g} \right)$ is the gradient of the {activated} constraints 
acquired in $\mathbf{g}=\left[\mathbf{g}_{\mathsf{S}}, \,\, \mathbf{g}_{\mathsf{O}} \right]$, where 
\begin{align}
&\mathbf{g}_{\mathsf{S}}=\mathbf{d}_{\mathsf{S}}-d_{\mathsf{S}}, \quad\,\, \mathbf{d}_{\mathsf{S}}=\left\{\dij: \dij< d_{\mathsf{S}}\right\}, \\
&\mathbf{g}_{\mathsf{O}}=\mathbf{d}_{\mathsf{O}}-d_{\mathsf{O}}, \quad \mathbf{d}_{\mathsf{O}}=\left\{l_{i,\mathsf{o}_j}^{(k)}: l_{i,\mathsf{o}_j}^{(k)}< d_{\mathsf{O}}\right\},
\end{align}
with $l_{i,\mathsf{o}_j}^{(k)}$ being the minimum distance of the $i$th \ac{UAV} from the $j$th obstacle, indicated as $\mathsf{o}_j$.
\begin{figure}[t!]
\psfrag{U}[lc][lc][0.8]{UAVs}
\psfrag{T}[lc][lc][0.8]{User}
\psfrag{B}[c][c][0.8]{Baseline}
\psfrag{x}[c][c][0.8]{x [m]}
\psfrag{y}[c][c][0.8]{y [m]}
    \centering
\includegraphics[width=0.5\textwidth]{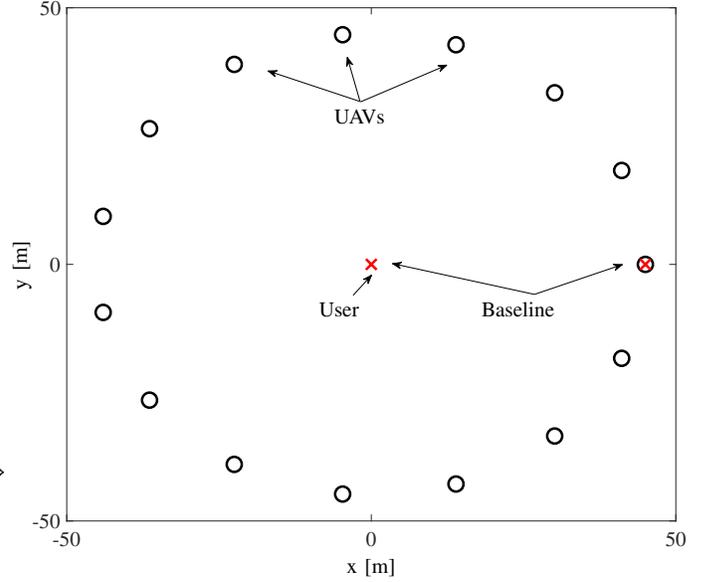}
\caption{Infrastructure-less localization scenario.}
\label{fig:loc_scenario}
\end{figure}

\section{Case Study}
\label{sec:cs}
\subsection{Infrastructure-less localization results}
\label{sec:rel_peb}
\begin{figure}[t!]
\psfrag{Data2}[lc][lc][0.8]{Ranging-only, $\xi=0$}
\psfrag{Data11111111111111}[lc][lc][0.8]{Ranging-only, $\xi=1$}
\psfrag{Data3}[lc][lc][0.8]{Bearing-only}
\psfrag{x}[c][c][0.8]{x [m]}
\psfrag{y}[c][c][0.8]{y [m]}
    \centering
\includegraphics[width=0.5\textwidth]{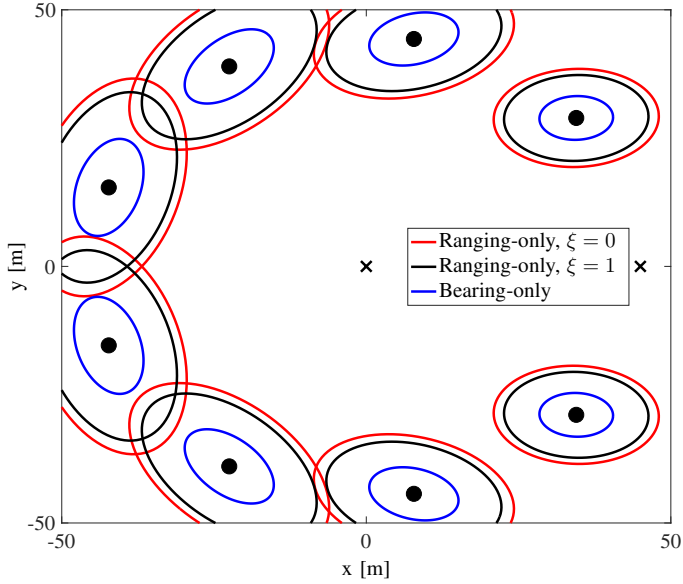} 
\caption{Localization error ellipses for different measurement models.}
\label{fig:Ellipses}
\end{figure}
\begin{figure}[t!]
\psfrag{Data2}[lc][lc][0.8]{Ranging-only, $\xi=0$}
\psfrag{Data11111111111111}[lc][lc][0.8]{Ranging-only, $\xi=1$}
\psfrag{Data3}[lc][lc][0.8]{Bearing-only}
\psfrag{RMSE}[c][c][0.8]{$\mathsf{RMSE}$ [m]}
\psfrag{N}[c][c][0.8]{Number of UAVs, $N$}
\psfrag{Time}[c][c][0.8]{Discrete time instant, $k$}
\psfrag{y}[c][c][0.8]{y [m]}
    \centering
\includegraphics[width=0.5\textwidth]{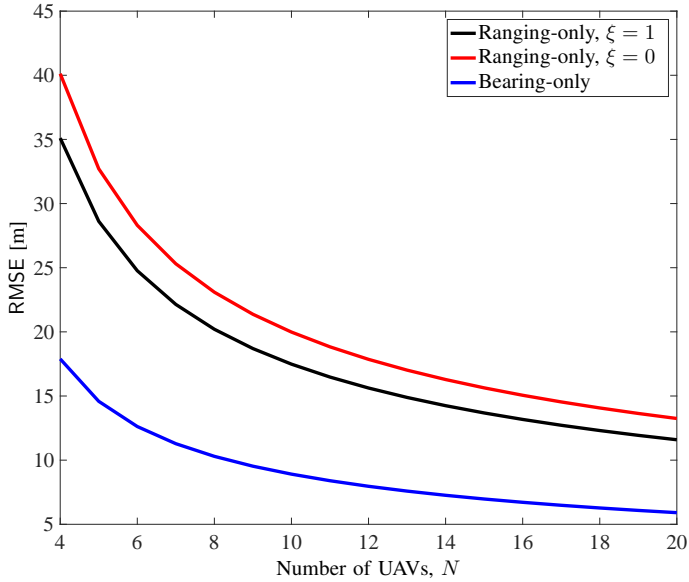}  
\caption{\ac{RMSE} as a function of the number of \acp{UAV}.}
\label{fig:RMSE_N}
\end{figure}
In this section, we report some results on the relative localization performance. We consider that the network of \acp{UAV} is distributed in a circumference centered at the user location, i.e., $\left[0,0 \right]^\mathsf{T}$, and with a radius of $45$ meters. The $x$-axis was defined by the location of the user and the actual position of the first \ac{UAV} which is $\left[45, 0 \right]^\mathsf{T}$. The positions of these two UAVs are considered known and they are indicated with  crosses in Figs.~\ref{fig:loc_scenario}-\ref{fig:Ellipses}. 
All the other \acp{UAV} are in unknown positions and they are represented by circles. The measurement standard deviations are set to $\sdo=0.39$ in logarithmic scale (or, equivalently, $\sigma_{\mathsf{sh}}/\alpha=1.7$ with $\sigma_{\mathsf{sh}}=3.4\,$dB) \cite{patwari2005locating} and $\sbo=10^\circ$ \cite{petitjean2018pimrc,isaacs2014quadrotor,isaacs2014gps}. 
 
We first calculated the \ac{CRLB} matrix $\bm{\kappa}$ by inverting \eqref{eq:classicalFIM} whose sub-\ac{FIM} elements are expressed in \eqref{eq:FIMe}. Then we computed the ellipses of the localization uncertainty starting from the \ac{CRLB} matrix associated to each UAV. For example, for the $i$th UAV, we have 
\begin{equation}\label{eq:CRBi}
\bm{\kappa}_i=\left[\begin{array}{cc}
\left[\bm{\kappa}_{\mathsf{xx}}\left(\bm{\theta} \right)\right]_{ii} & \left[\bm{\kappa}_{\mathsf{xy}}\left(\bm{\theta} \right)\right]_{ii}  \\
\left[\bm{\kappa}_{\mathsf{yx}}\left(\bm{\theta} \right)\right]_{ii} & \left[\bm{\kappa}_{\mathsf{yy}}\left(\bm{\theta} \right)\right]_{ii} 
\end{array}
 \right].
\end{equation}
The directions of the major axes of the ellipses are the eigenvectors of the matrix in \eqref{eq:CRBi}, whereas their lengths  correspond to their eigenvalues.

Figure~\ref{fig:Ellipses} presents the results for the following three cases: the red curves refer to ranging observations and ignorance of how the ranging model depends on the state vector, i.e., $\xi=0$; the black curves refer to ranging observations with a perfect knowledge of the observation model; i.e. $\xi=1$; and the blue curves are for the bearing case.  As expected, in the case of model awareness, the achievable localization results are better in comparison to the case where the position dependence in the ranging variance is not taken into account by the \ac{CRLB} computation.

Figure~\ref{fig:RMSE_N} displays the results in terms of \ac{RMSE} as a function of the number of \acp{UAV}. The \ac{RMSE} was obtained according to
\begin{equation}
\mathsf{RMSE}=\sqrt{\frac{\text{tr}\left(\bm{\kappa}_{\mathsf{xx}}\left(\bm{\theta} \right)+\bm{\kappa}_{\mathsf{yy}}\left(\bm{\theta} \right)\right)}{N_{\mathsf{u}}}},
\end{equation}
where $\text{tr}\left(\cdot \right)$ is the trace operator. It is not surprising that by increasing the number of \acp{UAV}, due to acquisition of more information, we obtain more accurate position estimates of the UAVs.
\begin{figure}[t!]
\psfrag{y}[c][c][0.65]{y [m]}
\psfrag{x}[c][c][0.65]{x [m]}
\psfrag{U}[lc][lc][0.35]{UAV}
\psfrag{t}[lc][lc][0.35]{trajectories}
\psfrag{Init}[lc][lc][0.35]{Initial UAV}
\psfrag{D}[lc][lc][0.35]{Desired UAV}
\psfrag{p}[lc][lc][0.35]{positions}
    \centering
\includegraphics[width=0.2\textwidth]{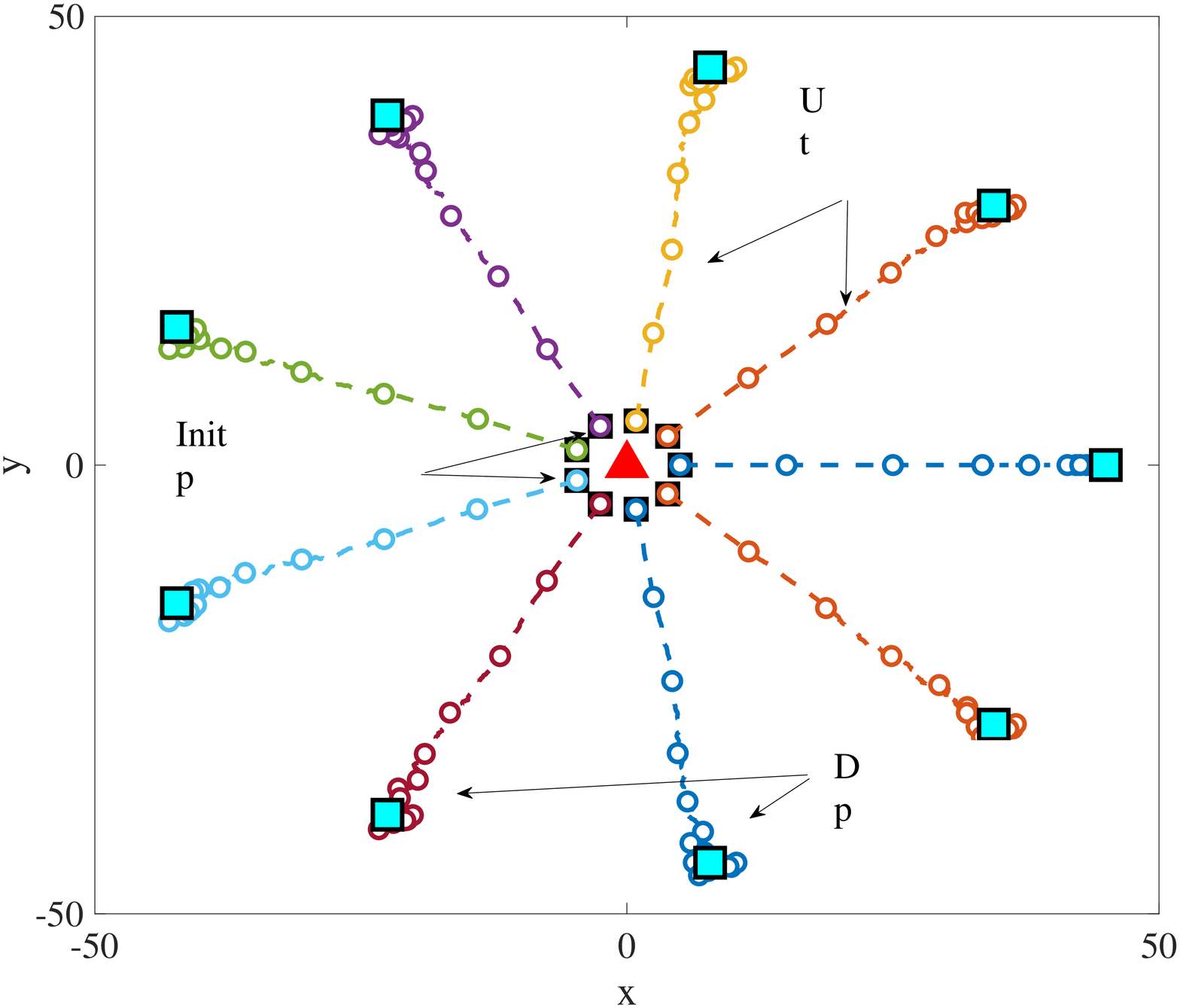}\qquad
 \includegraphics[width=0.2\textwidth]{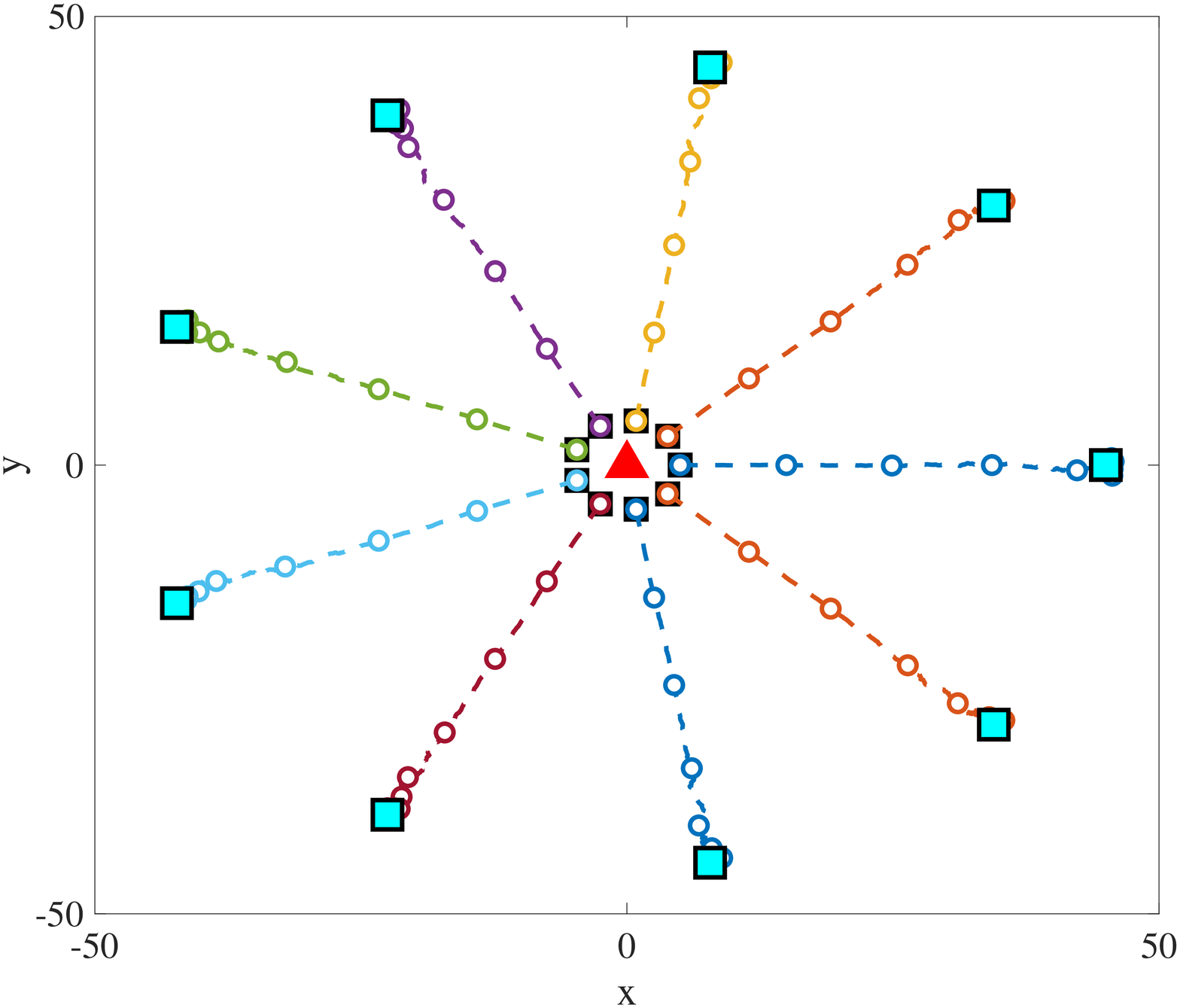}\\
 \vspace{0.5cm}
\includegraphics[width=0.2\textwidth]{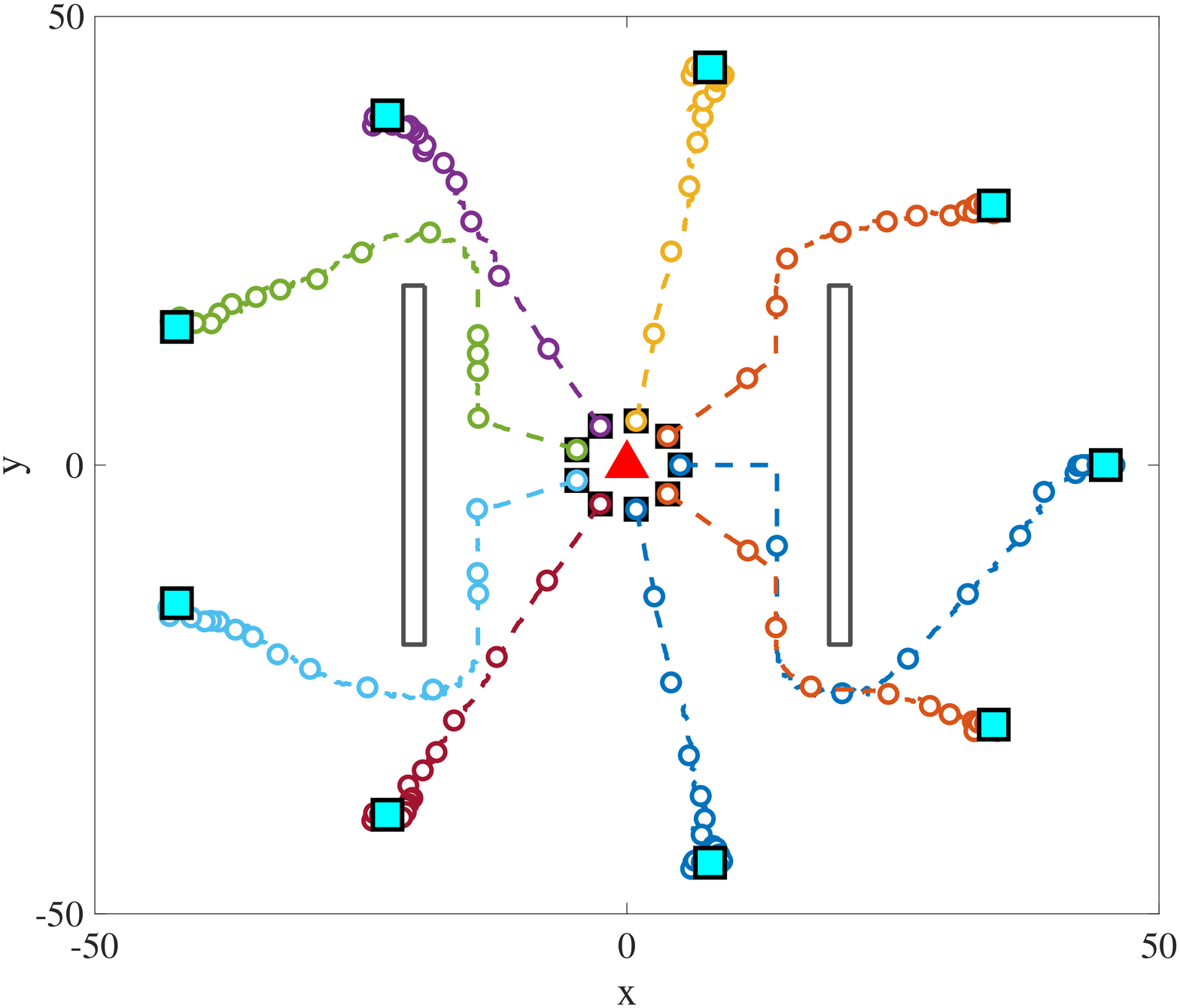}\qquad
 \includegraphics[width=0.2\textwidth]{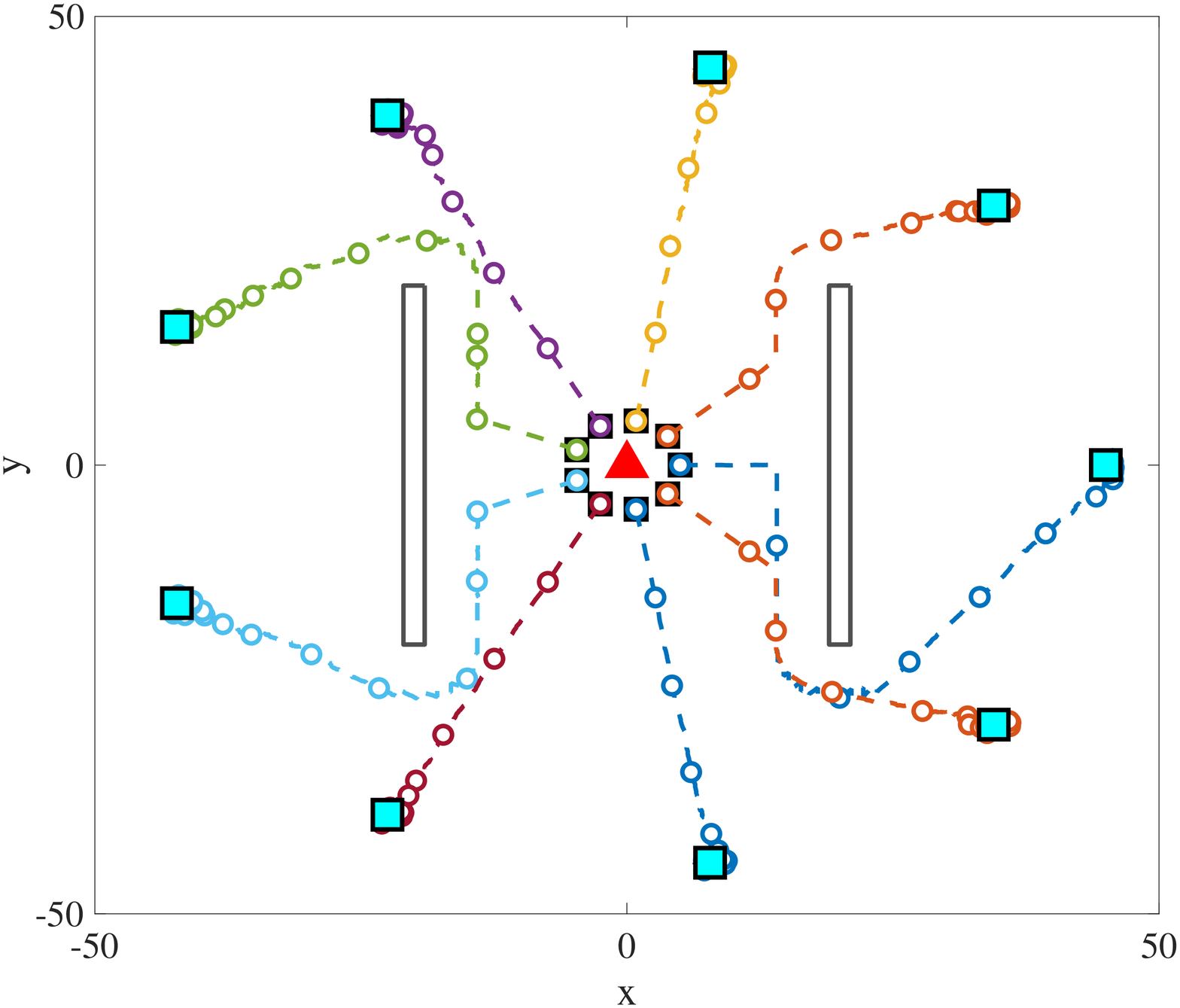}
 \caption{Example of navigation simulation. Top: \ac{LOS} propagation scenario; Bottom: \ac{NLOS} propagation scenario; Left: Ranging-only measurements; Right: Bearing-only measurements.}
 \label{fig:Form_scenario}
\end{figure}

The localization results discussed in this section were used as input for processing the navigation commands. In the following, we show the  achievable  accuracy in network formation when the relative \ac{CRLB} is used as the variance of the \ac{UAV} position estimates as indicator of the best performance achievable by any practical estimator.

\subsection{Network formation results}
\label{sec:nav}
In this section, we report the performance of the network formation task. More specifically, we suppose that at each time instant $k$, the positions of the \acp{UAV} are estimated and the variance is fixed to the \ac{CRLB} value at that  time instant. Then, the control signals are computed as the gradient of the distance between the estimated and the desired final UAV positions. Navigation constraints are also included in the simulation, with $d_{\mathsf{S}}=0.5\,$m, $d_{\mathsf{O}}=5\,$m, and $v=1\,$m/s.
\begin{figure}[t!]
\psfrag{Data111111111111111}[lc][lc][0.8]{Ranging-only, LOS}
\psfrag{Data2}[lc][lc][0.8]{Bearing-only, LOS}
\psfrag{Data3}[lc][lc][0.8]{Ranging-only, NLOS}
\psfrag{Data4}[lc][lc][0.8]{Bearing-only, NLOS}
\psfrag{RMSE}[c][c][0.8]{$\mathsf{RMSE}^{(k)}$ [m]}
\psfrag{Time}[c][c][0.8]{Discrete time instant, $k$}
    \centering
\includegraphics[width=0.5\textwidth]{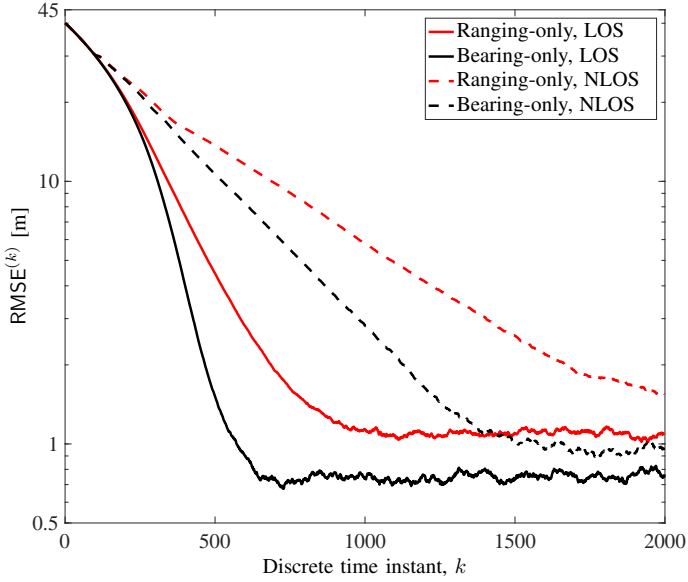} 
\caption{\ac{RMSE} as a function of time.}
\label{fig:RMSE_formation}
\end{figure}

\begin{figure}[t!]
\psfrag{RMSE}[c][c][0.8]{$\mathsf{RMSE}^{(k)}$ [m]}
\psfrag{Time}[c][c][0.8]{Discrete time instant, $k$}
\psfrag{L}[lc][lc][0.8]{LOS}
\psfrag{N}[lc][lc][0.8]{NLOS}
\psfrag{Data11111111}[lc][lc][0.8]{$\sigma_{\mathsf{sh}} \!=\! 0.5$ dB} 
\psfrag{Data2}[lc][lc][0.8]{$\sigma_{\mathsf{sh}} \!=\! 1.5$ dB} 
\psfrag{Data3}[lc][lc][0.8]{$\sigma_{\mathsf{sh}}\!=\! 2.5$ dB} 
\psfrag{Data4}[lc][lc][0.8]{$\sigma_{\mathsf{sh}} \!=\! 3.5$ dB} 
\psfrag{Data5}[lc][lc][0.8]{$\sigma_{\mathsf{sh}} \!=\! 4.5$ dB} 
\psfrag{Data6}[lc][lc][0.8]{$\sigma_{\mathsf{sh}} \!=\! 5.5$ dB} 
\includegraphics[width=0.5\textwidth]{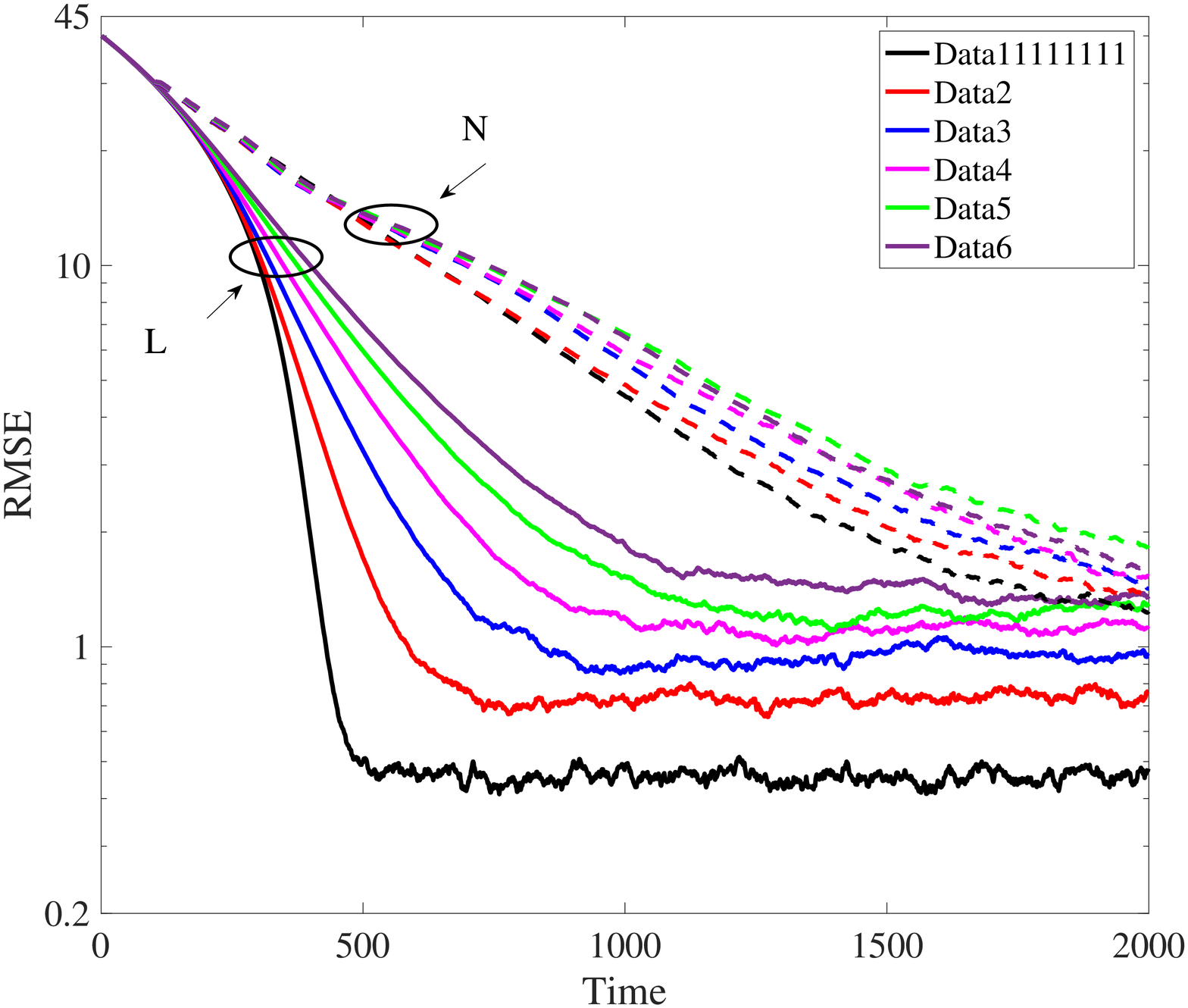} 
\psfrag{Data111111}[lc][lc][0.8]{$\sbo=5^\circ$}
\psfrag{Data2}[lc][lc][0.8]{$\sbo=10^\circ$}
\psfrag{Data3}[lc][lc][0.8]{$\sbo=20^\circ$}
\psfrag{Data4}[lc][lc][0.8]{$\sbo=30^\circ$}
\includegraphics[width=0.5\textwidth]{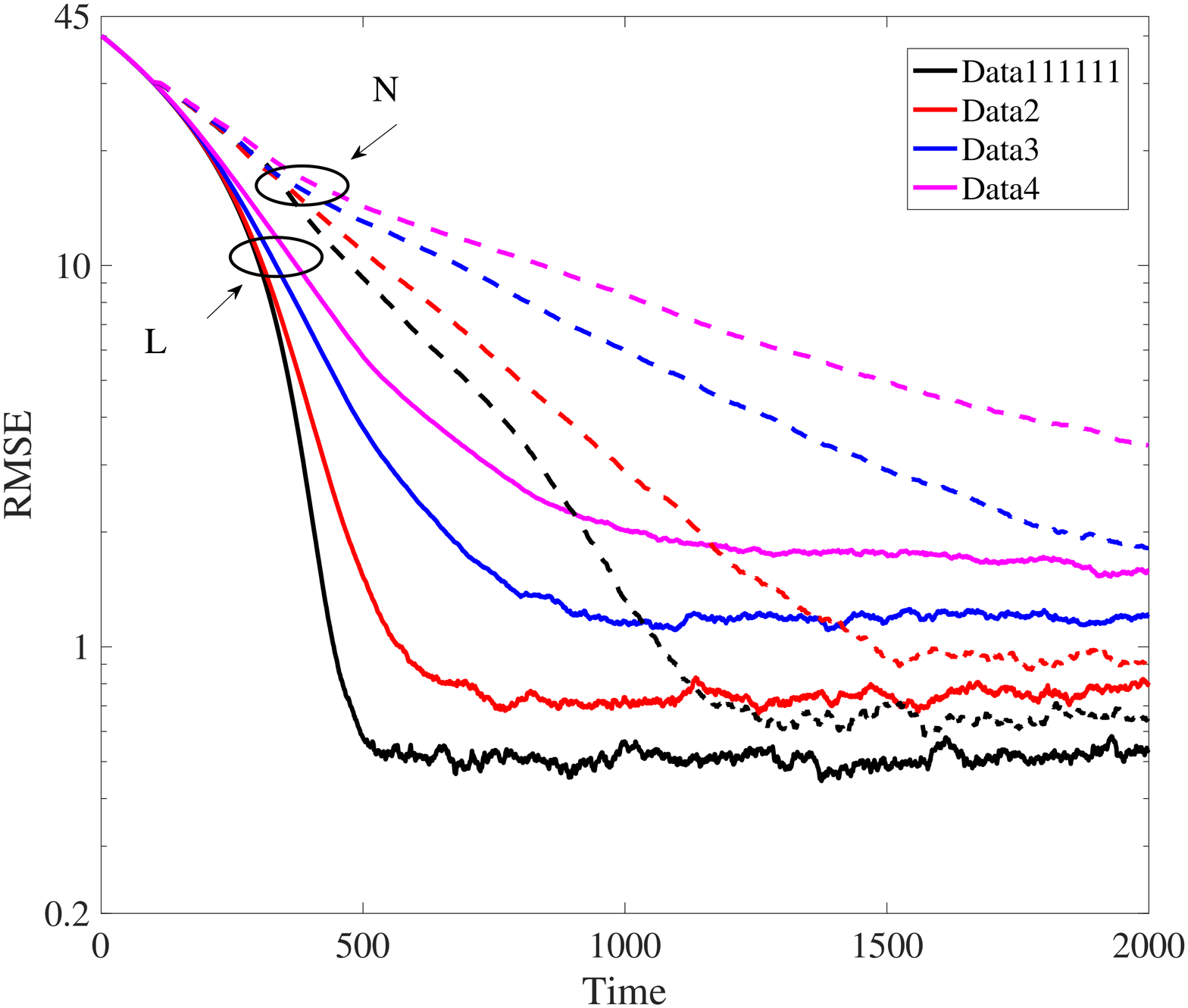}
\caption{\ac{RMSE} as a function of the adopted sensor accuracy. Top: ranging-only measurements; Bottom: bearing-only measurements.}
\label{fig:RMSE_technology}
\end{figure}
The simulated scenarios are reported in Fig.~\ref{fig:Form_scenario} for a circular shape formation in \ac{LOS} and \ac{NLOS} environments and for ranging and bearing measurements. In the plot, the user is indicated with a red triangle, the \ac{UAV} initial and final positions with white and cyan squares, respectively, and the \ac{UAV} trajectories with dashed lines. Also some intermediate \ac{UAV} positions along the trajectories are displayed every $100$ time slots with the aim of clarifying that the superimposed trajectories do not coincide with collision events between the UAVs. The obstacles are depicted as grey rectangles. The performance is  investigated in terms of \ac{RMSE} averaged over the number of \acp{UAV}, i.e.,
\begin{equation}
\mathsf{RMSE}^{(k)}=\frac{1}{N} \sum_{i=1}^{N}\,\sqrt{\frac{1}{N_{\mathsf{MC}}}\, \sum_{m=1}^{N_{\mathsf{MC}}} \lVert \bm{\theta}_{im}^{(k)} - \bm{\theta}_i^{*}  \rVert^{{2}} },
\end{equation}
with $N_{\mathsf{MC}}$ being the number of Monte Carlo trials, where at each iteration a different \ac{UAV} position estimate is generated. In Fig.~\ref{fig:RMSE_formation}, the red and black curves are for ranging-only and bearing-only measurements, respectively. The dashed lines refer to the presence of obstacles in the simulated environment. The number of \acp{UAV} is fixed to $N=9$, the number of Monte Carlo trials to $N_{\mathsf{MC}}=100$, and the multipath bias to $\sigma_\mathsf{b}=3\,$dB. We notice that at the last time instant, the \acp{UAV} attain the final desired formation with an average error of less than $2$ m. As expected, the \ac{NLOS} case decreases the performance, especially for the bearing case. A comparison of results based on ranging and bearing measurements is not meaningful because the performance strongly depends on the accuracy of the sensors that acquire the measurements. 

In order to gain insights about the dependence of the performance of the proposed method on this accuracy, we investigated the RMSE of the UAV positions as a function of the sensor accuracy. More specifically, in   Fig.~\ref{fig:RMSE_technology}, we provide the \ac{RMSE} performance for different choices of the parameters, i.e., $\sigma_{\mathsf{sh}}$, (we suppose that the path loss coefficient is always equal to $2$) and $\sbo$. As expected, from the results of the ranging case, we observe that the \ac{RSS}-based distance estimation and, consequently, the positioning become more accurate when the standard deviation of the shadowing decreases. For the bearing case, $\sbo=5^\circ$ is an accuracy achievable with the adoption of antenna arrays whose integration in drones is today difficult due to size and weight constraints. Also in this case, relying on \ac{RSS} measurements can be a solution at the expense of a degradation in localization and formation performance.

\section{Conclusions}
\label{sec:conclusions}
In this paper, we investigated the performance of a user-centric network of \acp{UAV} in formation and infrastructure-less 
localization in indoor environments. The \acp{UAV} are able to exchange measurements with all the other UAVs of the network and communicate the collected observations to a central node, i.e., the user. 
From ranging and bearing observation models, the user is able to localize the \acp{UAV} and to send them navigation commands with the objective of forming a  desired final topology. The results of formation accuracy are promising and pave the way to indoor applications using \acp{UAV}. In future work, we will consider scenarios with dynamic users.  

\section*{Acknowledgment}
This work has received funding from the European Union's Horizon 2020 research and innovation programme under the Marie Sklodowska-Curie project AirSens (grant no. 793581) and the support of the NSF under Award CCF-1618999. 

\bibliographystyle{IEEEtran}

\end{document}